\documentclass[usenatbib,useAMS]{mn2e}
\usepackage{times}
\usepackage{epsfig}

\begin{document}
\title[Planetary mass function and planetary systems]{Planetary mass function and planetary systems}
\author[M. Dominik]{M. Dominik,$^{1}$\thanks{Royal Society University Research Fellow}\thanks{E-mail: md35@st-andrews.ac.uk}\\
$^{1}$SUPA, University of St Andrews, School of Physics \& Astronomy, 
North Haugh, St Andrews, KY16 9SS, United Kingdom}

\maketitle

\begin{abstract}
With planets orbiting stars, a planetary mass function should not be seen as a low-mass extension of the stellar mass function, but a proper formalism needs to take care of the fact that the statistical properties of planet populations are linked to the properties of their respective host stars. This can be accounted for by describing planet populations by means of a differential planetary mass-radius-orbit function, which together with the fraction of stars with given properties that are orbited by planets and the stellar mass function allows to derive all statistics for any considered sample. These fundamental functions provide a framework for comparing statistics that result from different observing techniques and campaigns which all have their very specific selection procedures and detection efficiencies. Moreover, recent results both from gravitational microlensing campaigns and radial-velocity surveys of stars indicate that planets tend to cluster in systems rather than being the lonely child of their respective parent star. While planetary multiplicity in an observed system becomes obvious with the detection of several planets, its quantitative assessment however comes with the challenge to exclude the presence of further planets. Current exoplanet samples begin to give us first hints at the population statistics, whereas pictures of planet parameter space in its full complexity call for samples that are 2--4 orders of magnitude larger. In order to derive meaningful statistics however, planet detection campaigns need to be designed in such a way that well-defined fully-deterministic target selection, monitoring, and detection criteria are applied. The probabilistic nature of gravitational microlensing makes this technique an illustrative example of all the encountered challenges and uncertainties.
\end{abstract}


\section{Introduction}

More than 450 planets orbiting stars other than the Sun have been detected to date by means of four different techniques: Doppler-wobble stellar radial-velocity measurements, planetary transits, gravitational microlensing, and the direct detection of emitted or reflected light.
Observing campaigns now need to evolve from the pure detection of planets 
to studies that allow to infer the statistical properties of the underlying populations that are being probed. In order to achieve such a goal, 
deterministic procedures for the selection of targets and the identification of planetary signals are required \citep{SIGNALMEN,ARTEMiS,RoboNet-II,OT1,MiNDSTEp2008}.

While many studies on planet populations based on 
data from radial-velocity surveys have been carried out \citep[e.g.\ ][]{Marcy:abundance1,Udry:abundance,Marcy:abundance2,Mayor:abundance,OT2},
there has been a long silence on extracting planetary abundances from gravitational microlensing campaigns since the twin papers on
the first five years of the PLANET campaign\footnote{{\tt http://www.planet-legacy.org}} \citep{PLANET:fiveshort,PLANET:fivelong}, and those on the 
OGLE-II  \citep{OGLE2:limits} and the first year of the OGLE-III
survey\footnote{{\tt http://ogle.astrouw.edu.pl}} \citep{OGLE2002}.
Only very recently, there appears to be a sudden inflation \citep{Sumi:planet,Gould:abundance}.

The various techniques currently used for studying planet populations (and any future ones as well) have their very own preferred regions of planet
parameter space that they are sensitive to, while being blind to others.
Consequently, these do not directly probe {\em the} planetary mass function, but some bits and pieces for which the detection efficiency of the campaign and the selection biases need to be determined carefully. Any quoted
planet abundance needs to come with a complete description which region
of planet parameter space it refers to, and how it has been averaged.
In particular, one can anticipate substantial differences on whether
one talks about Solar-type stars or M dwarfs, hot or cool planets, bulge or disk stars (with their different metallicities). 


In order to get around such difficulties, and ease the comparison between findings that arise from different campaigns and/or techniques, a general
framework of differential planetary mass functions with respect to their
fundamental parameters is suggested in this paper. 

Rather than planets being distributed just randomly amongst stars, it seems 
that they tend to cluster in planetary systems. Such a conjecture is underpinned by recent observational evidence on both outer gas-giant planets \citep{DoubleCatch,Marois:planet} as well as Super-Earths and planets with Neptune-class masses in closer orbits \citep{Mayor:abundance,HARPS:abundance2}. Therefore, the detection of planets in an experiment does not correspond to independent draws from its population, but the probability of their detection around a star that is known to host planets is larger than that of finding it around a randomly chosen star.

While Sect.~\ref{sec:formalism} presents a theoretical framework
for describing planet populations in view of clustering in planetary
systems and specific regions of interest or sensitivity,
Sect.~\ref{sec:measureMF} provides rough estimates for the size of planet samples required to assess the fundamental functions that decribe these populations. Sect.~\ref{sec:multiplicity} is devoted to planetary multiplicity, while Sect.~\ref{sec:abundance} looks into planet abundance estimates arising from gravitational microlensing observations, their uncertainties, and the involved challenges. Sect.~\ref{sec:outlook} finally concludes the paper with a short summary and outlook.

\section{Fundamental functions describing planetary systems}
\label{sec:formalism}

Only celestial bodies that are in orbit around a around a
star or stellar remnant are `planets'.\footnote{following the ``Position Statement  on the Definition of a `Planet'{''} (in the revision dated 28 February 2003) by the IAU Working Group on Extrasolar Planets (WGESP)} This means that planets
cannot be seen in isolation from these, and consequently
planets are not well-described by just extending the stellar mass function \citep{Salpeter,Scalo:IMFreview,Kroupa:Science} to lower masses, but a mass function decribing planets
needs to link to their host stars (or remnants).

Let us consider explicitly the dependence of planetary abundance on stellar mass $M_\star$, metallicity $Z$, age $\tau$, and spin rate $\Omega$ and therefore
define a differential stellar mass function 
$\xi(M_\star,Z,\tau,\Omega)$,\footnote{For the vast majority of stars, the spin rate $\Omega$ essentially becomes a function of stellar mass $M_\star$ and age $\tau$ \citep[e.g.\ ][]{CCLi,Barnes}, so that this parameter can be neglected.} so that 
for a population with density functions $p_Z(Z)$, $p_\tau(\tau)$,
and $p_\Omega(\Omega)$ for metallicity, age, or spin rate, respectively,
one obtains a mass function
\begin{equation}
\Xi(M_\star) = \int \xi(M_\star,Z,\tau,\Omega)\,p_Z(Z)\,\mathrm{d}Z\,p_\tau(\tau)\,\mathrm{d}\tau\,p_\Omega(\Omega)\,\mathrm{d}\Omega\,,
\end{equation}
where the number density of stars becomes
\begin{equation}
N_\star = \int\,\Xi(M_\star)\,\rmn{d}[\lg(M_\star/M_\odot)]\,,
\end{equation}
where $M_{\odot}$ denotes the mass of the Sun.

For the stars that host planets, the properties of planets can then be
described by a differential planetary mass-radius-orbit function
$\varphi(m_\rmn{p},r_\rmn{p},a,\varepsilon; M_\star, Z, \tau,\Omega)$, where $m_p$, $r_\rmn{p}$, $a$, and $\varepsilon$ denote the mass, radius, orbital semi-major axis, or orbital eccentricity of the planet, respectively, and further parameters might be added. This implies a mass function for
planetary systems around stars with $(M_\star,Z,\tau,\Omega)$ given by
\begin{eqnarray}
& & \hspace*{-2em}
\Phi(m_\rmn{p};M_\star, Z, \tau,\Omega) = \int
 \varphi(m_\rmn{p},r_\rmn{p},a,\varepsilon;M_\star,Z,\tau,\Omega)\;\times
\nonumber \\
& & \times\;
\rmn{d}[\lg(r_\rmn{p}/r_\oplus)]\,\rmn{d}[\lg(a/a_\oplus)]\,\rmn{d}\varepsilon\,,
\end{eqnarray}
with $r_\oplus$ being the Earth's radius and $a_\oplus = 1~\mbox{au}$,
so that the average number of planets in such systems reads
\begin{equation}
n_\mathrm{p}(M_\star, Z, \tau,\Omega) = \int \Phi(m_\rmn{p};M_\star, Z, \tau,\Omega)
\,\rmn{d}[\lg(m_\rmn{p}/M_\oplus)]\,,
\end{equation}
where $M_\oplus$ is 
the mass of the Earth.

With $f_\mathrm{p}(M_\star,Z,\tau,\Omega)$ denoting the fraction of stars that 
host planets, the number density of planets for a stellar population
becomes
\begin{eqnarray}
& & \hspace*{-2em} 
N_\mathrm{p} = \int f_\mathrm{p}(M_\star,Z,\tau,\Omega)\;
\xi(M_\star,Z,\tau,\Omega)\;
n_\mathrm{p}(M_\star,Z,\tau,\Omega)\;\times \nonumber \\
& & \times\;p_Z(Z)\,
\mathrm{d}Z\,p_\tau(\tau)\,
\mathrm{d}\tau\,p_\Omega(\Omega)\,
\mathrm{d}\Omega\,\mathrm{d}[\lg(M/M_\star)]\,.
\end{eqnarray}

Moreover, one finds a population-integrated planetary mass-radius-orbit function 
\begin{eqnarray}
& & \hspace*{-2em} \psi(m_\rmn{p},r_\rmn{p},a,\varepsilon) = \int f_\rmn{p}(M_\star,Z,\tau,\Omega)\;\xi(M_\star,Z,\tau,\Omega)\;\times \nonumber\\
& & \times\;\varphi(m_\rmn{p},r_\rmn{p},a,\varepsilon; M_\star, Z, \tau,\Omega)\;\times \nonumber \\
& & \times\;\,p_Z(Z)\,\rmn{d}Z\,p_\tau(\tau)\,
\rmn{d}\tau\,p_\Omega(\Omega)\,
\rmn{d}\Omega\,\rmn{d}[\lg(M_\star/M_\odot)] \,,
\end{eqnarray} 
and a corresponding planetary mass function results as
\begin{equation}
\Psi(m_\rmn{p}) = \int \psi(m_\rmn{p},r_\rmn{p},a,\varepsilon)\,
\rmn{d}[\lg(r_\rmn{p}/r_\oplus)]\,\rmn{d}[\lg(a/a_\oplus)]\,\rmn{d}\varepsilon\,,
\end{equation}
so that one finds the number density of planets again as
\begin{equation}
N_\mathrm{p} = \int \Psi(m_\rmn{p})\,\rmn{d}[\lg(m_\rmn{p}/M_\oplus)]\,.
\end{equation}

Provided that experiments in the hunt for extra-solar planets follow deterministic criteria, 
a mass function can be extracted that refers to the selected host stars 
and planetary orbits that the applied technique is sensitive to,
i.e.\ averages are taken over the stellar population and the 
orbital parameters. However, in order to answer fundamental questions
such as `How frequent are planets of a given mass
range in the Solar neighbourhood?', `What fraction of stars in the
Milky Way do have planetary systems?', or `How many planets that could host life are there in the Universe?',
one needs to trace back the description of planetary systems 
to more fundamental functions such as the differential mass-radius-orbit function
$\varphi(m_\rmn{p},r_\rmn{p},a,\varepsilon; M_\star, Z, \tau,\Omega)$, the fraction of
stars with planetary systems $f_\mathrm{p}(M_\star, Z, \tau,\Omega)$,
and the differential stellar mass function $\xi(M_\star,Z,\tau,\Omega)$.

\section{Measuring planetary mass functions}
\label{sec:measureMF}

In order to obtain an estimate on how well we can measure a planetary mass function, let us consider dividing the parameter space into multi-dimensional bins. If rather than aiming for a precision measurement, one sets the goal at an 'astronomical' accuracy of 50 per cent, the assumption
of Poisson statistics yields the requirement of each bin to contain at least 4 planets. Let $p$ denote the number of considered parameters, and $b$ the number of considered parameter ranges, the minimal number of planets needed to provide the desired result is $N_\rmn{p} = 4\,b^{p}$, which would correspond to letting the choice of parameter ranges follow the observed distribution of detected planets, in such a way that each bin contains exactly the mimimum of 4 planets. With $\kappa$ denoting the desired accuracy, one finds more generally $N_\rmn{p} = \kappa^{-1/2}\,b^{p}$.
Table~\ref{tab:nplanets} shows the requirements for some selected cases with relative accuracies of 50 per cent or 20 per cent, 2-, 4- or 6-parameter functions, and a various number of bins ranging from 2 to 10. Roughly, one gets an idea of the distribution of the planet abundances with $b \geq 3$, but
one can realistically only start talking about a ``planetary mass function'' for $b \geq 5$. While a planetary mass-radius-separation function $\varphi_{m_\rmn{p},r_\rmn{p},a}(m_\rmn{p},r_\rmn{p},a;M_\star,Z,\tau)$
depending on the stellar mass, metallicity, and age involves 6 parameters,
less detailed 4-parameter functions are e.g. the
planetary mass-separation  $\varphi_{m_\rmn{p},a}(m_\rmn{p},a;M_\star,Z)$ or mass radius function  $\varphi_{m_\rmn{p},r_\rmn{p}}(m_\rmn{p},r_\rmn{p};M_\star,Z)$ depending on stellar mass and
metallicity, or a planetary mass-radius-separation function depending
on stellar mass only, and 2-parameter functions would e.g. be 
the planetary mass function $\varphi_{m_\rmn{p}}(m_\rmn{p};M_\star)$ depending on stellar mass only, or the
planetary mass-separation function $\varphi_{m_\rmn{p},a}(m_\rmn{p},a)$ irrespective of the stellar properties.
We now have a total sample of about 450 planets orbiting stars other than the Sun, where it took about 10~years to detect the first~150, then about 3~years to detect the next~150, and then just about 1~year to detect
the equal number of~150. Table~\ref{tab:nplanets} shows how long campaigns
with a constant detection rate of 150 planets per year would have to last in order to obtain the respective functions with desired 
accuracies.

Right now, the collected data allow to measure 1-parameter functions, find the basic structure structure ($b \geq 10$) of 2-parameter functions, see basic trends ($b \geq 3$) in 4-parameter functions, and some hint on the dependency of the planet abundance on further parameters. With 150 planets per year, or more realistically, a fair factor of this rate, rough ideas ($b \geq 5$) of 4-parameter planetary mass functions ($b \geq 5$) and an indication of trends ($b \geq 3$) for 6-parameter planetary mass functions are obtainable within foreseeable time frames, but the numbers call for more aggressive searches.

\begin{table}
\begin{tabular}{ccclc}
\hline
$\kappa$ & $p$ & $b$ & $N_\rmn{p}$ & $T_{150}$\\
\hline
0.5 & 2 & 2 & $4\times 2^2 = 16$ & 1.3~months \\
0.5 & 2 & 3 & $4\times 3^2 = 36$ & 2.9~months \\
0.5 & 2 & 5 & $4\times 5^2 = 100$ & 8~months \\
0.5 & 2 & 10 & $4\times 10^2 = 400$ & 3~years \\ \hline
0.5 & 4 & 2 & $4\times 2^4 = 64$ & 5~months \\
0.5 & 4 & 3 & $4\times 3^4 = 324$ & 2~years \\
0.5 & 4 & 5 & $4\times 5^4 = 2500$ & 17~years \\
0.5 & 4 & 10 & $4\times 10^4 = 40,000$ & 270~years \\ \hline
0.5 & 6 & 2 & $4\times 2^6 = 256$ & 1.7~years \\
0.5 & 6 & 3 & $4\times 3^6 = 2916$ & 19~years \\
0.5 & 6 & 5 & $4\times 5^6 = 62,500$ & 420~years \\
0.5 & 6 & 10 & $4\times 10^6 = 4,000,000$ & 27,000~years \\ \hline
0.2 & 2 & 2 & $25\times 2^2 = 100$ & 8~months\\
0.2 & 2 & 3 & $25\times 3^2 = 225$ & 1.5~years \\
0.2 & 2 & 5 & $25\times 5^2 = 625$ & 4~years \\
0.2 & 2 & 10 & $25\times 10^2 = 2500$ & 17~years \\ \hline
0.2 & 4 & 2 & $25\times 2^4 = 400$ & 2.7~years\\
0.2 & 4 & 3 & $25\times 3^4 = 2025$ & 13.5~years \\
0.2 & 4 & 5 & $25\times 5^4 = 15,625$ & 100~years \\
0.2 & 4 & 10 & $25\times 10^4 = 250,000$ & 1700~years \\ \hline
0.2 & 6 & 2 & $25\times 2^6 = 1600$ & 11~years \\
0.2 & 6 & 3 & $25\times 3^6 = 18,225$ & 120~years \\
0.2 & 6 & 5 & $25\times 5^6 = 390,625$ & 2600~years \\
0.2 & 6 & 10 & $25\times 10^6 = 25,000,000$ & 170,000~years \\ \hline
\end{tabular}
\caption{Minimal number of planets $N_\rmn{p}$ required to sample a descriptive
statistic with $p$ parameters with $b$ bins to a relative accuracy $\kappa$, and
time $T_{150}$ required to acquire such a sample for a planet detection rate
of 150 per year.}
\label{tab:nplanets}
\end{table}

\section{Planetary multiplicity}
\label{sec:multiplicity}
While stars with and without planets have been distinguished by referring
to the fraction $f_\rmn{p}(M_\star,Z,\tau,\Omega)$ of stars that host planets and defining the differential planetary mass-radius-orbit function
$\varphi(m_\rmn{p},r_\rmn{p},a,\varepsilon; M_\star, Z, \tau,\Omega)$ to relate to these only, a further statistic is the distribution of the number of planets amongst all planetary systems. With multiplicity indices $\zeta_k$ that denote the fraction of planetary systems containing $k$ planets,
where 
\begin{equation}
 \sum_{k=1}^{\infty} \zeta_k = 1\,,
\end{equation}
the planetary mass-radius-orbit function can be decomposed as
\begin{eqnarray}
& & \hspace*{-2em}
\varphi(m_\rmn{p},r_\rmn{p},a,\varepsilon; M_\star, Z, \tau,\Omega)  \nonumber \\
& & =  \sum_{k=1}^{\infty} k\,\zeta_k\,\hat{\varphi}_k(m_\rmn{p},r_\rmn{p},a,\varepsilon; M_\star, Z, \tau,\Omega)\,,
\end{eqnarray}
where
\begin{eqnarray}
& & \hspace*{-2em}
\int \hat{\varphi_k}(m_\rmn{p},r_\rmn{p},a,\varepsilon; M_\star, Z, \tau,\Omega)\;\times \nonumber \\
& & \times\;\rmn{d}[\lg(m_\rmn{p}/M_\oplus)]\,\rmn{d}[\lg(r_\rmn{p}/r_\oplus)]\,\rmn{d}[\lg(a/a_\oplus)]\,\rmn{d}\varepsilon = k\,.
\end{eqnarray}

In general, all $\hat{\varphi}_k(m_\rmn{p},r_\rmn{p},a,\varepsilon; M_\star, Z, \tau,\Omega)$ may be different. Together with the multiplicity indices $\zeta_k$, one would be left with an infinite number of parameters. This however can be meaningfully avoided by adopting a functional dependence of $\zeta_k$ and $\hat{\varphi}_k$ on k that is described by a small finite number of parameters.

In particular, one might want to distinguish
stars with a single planets to multiple-planet systems, described by $\zeta_1$ (with $\zeta_\rmn{mult} = 1-\zeta_1$),
$\hat{\varphi}_1(m_\rmn{p},r_\rmn{p},a,\varepsilon; M_\star, Z, \tau,\Omega)$, and
\begin{eqnarray}
& & \hspace*{-2em}
\hat{\varphi}_\rmn{mult}(m_\rmn{p},r_\rmn{p},a,\varepsilon; M_\star, Z, \tau,\Omega)  \nonumber \\
& & =  \sum_{k=2}^{\infty} k\,\zeta_k\,\hat{\varphi}_k(m_\rmn{p},r_\rmn{p},a,\varepsilon; M_\star, Z, \tau,\Omega)\nonumber \\
& & = \varphi(m_\rmn{p},r_\rmn{p},a,\varepsilon; M_\star, Z, \tau,\Omega)\;- \nonumber \\
& & \hspace*{2em} -\;\zeta_1\,\hat{\varphi}_1(m_\rmn{p},r_\rmn{p},a,\varepsilon; M_\star, Z, \tau,\Omega)\,.
\end{eqnarray}
In fact, \citet{Wright} have argued that there is evidence for $\hat{\varphi}_1$ being different from $\hat{\varphi}_\rmn{mult}$.

The assessment of planetary multiplicity however poses a huge challenge for properly interpreting the observational data, given that our knowledge of the absence of further planets in observed systems is quite limited. If Hot Jupiters are considered lonely, whereas Neptune-mass planets are frequently found in multiple systems \citep{Mayor:abundance,HARPS:abundance2}, how much does this have to be attributed to the fact that observational techniques that report Hot Jupiters are insensitive to less massive planets, whereas if the sensitivity extends down to lower masses, other such planets are spotted rather easily? It is intriguing to see that observations of transit timing variations led to the suggestion of the presence of a 15~Earth-mass planet in the WASP-3 system \citep{MacPlanet} that was already known to host a Hot Jupiter \citep{WASP3}. Planets reported by microlensing in particular cannot be claimed to be the only ones in the system, they were just the only ones that revealed their presence during a transient event. \citet{390further} explicitly found that the acquired data do not exclude the presence of gas-giant planets at any separation orbiting the lens star that caused event OGLE-2005-BLG-390, which is known to host a cool Super-Earth \citep{PLANET:planet}. It is easier to detect a planet than being able to claim that there are no other planets orbiting the same star, and if one aims for quantifying multiplicity, this needs to be addressed.

Planetary multiplicity however becomes an obvious phenomenon with the detection of respective systems, such as 
the pair of gas-giant planets orbiting
OGLE-2006-BLG-109L \citep{DoubleCatch}, which resemble a half-scale version
of the Jupiter-Saturn part of the Solar system. Interestingly, the planets
found to orbit HR~8799 look like the complementary double-scale version
\citep{Marois:planet}. It is particularly striking that very early 
opportunities to detect such systems by gravitational microlensing or direct imaging, respectively, were successful, while one needs to keep in mind that planets with an orbital period similar to Saturn cannot be detected from radial-velocity surveys so far (given a 10--15 year history of respective campaigns), and observing planetary transits is
further disfavoured by the small transit probability.
However, for Super-Earths and planets with Neptune-class masses in closer orbits, radial-velocity surveys find a very high level of multiplicity 
as well \citep{Mayor:abundance,HARPS:abundance2}. 

The detection of the pair of Jupiter- and Saturn-like planets orbiting
OGLE-2006-BLG-109L \citep{DoubleCatch} is often hailed because of the striking similarity with the Solar system, albeit that there is basically nothing that can be said about potential inner rocky planets other than that such cannot be excluded. There is however another important result arising from this discovery: outer gas-giant planets are not of the lonesome type. How does one arrive at such a conclusion? Regardless
of the large detection efficiency for such planets in events with
a peak magnification as large as that of OGLE-2006-BLG-109 ($A_0 \sim 290$), the planetary abundance is moderate or small. If
we consider an abundance of 5 per cent, the probability for a double catch would be just 0.25 per cent if the planets were drawn independently from the population. This would mean an expected detection of $\sim\,1/30$
systems amongst the 13 events comprising the systematic sample 
reported by \citet{Gould:abundance}, so that we would have been very lucky to find the detected pair. Therefore, it appears 
the more likely assumption that the two detections were not the result
of independent draws, but instead the probability for a planet to orbit
a star is larger if one considers a star that is known to host planets
as compared to an arbitrarily chosen star that might host planets or not.
This however means that it is not appropriate to consider a planetary
mass function with planets randomly drawn from it, but instead one needs
to distinguish between stars with or without planets, as the formalism
suggested in the previous section does.
These arguments however get weaker if the planetary abundance was 
as large as 20 per cent, because this would mean a probability of
4 per cent for a pair, or 1/2 expected to be detected for 13 events
as compared to the one found.



\section{Planet abundance estimates from microlensing observations}
\label{sec:abundance}

The statistical analysis of microlensing events in order to derive planet abundance estimates
provides an illustrative example of the challenges one is facing. If a planet orbits the lens star,
a detectable signal will only arise with a finite probability. This finite detection efficiency for planets 
of given mass and orbital separation from their host star is of relevance not only for assessing abundances by means of detections
but also for drawing conclusions from the absence of planetary signals. Moreover, 
the host stars of planets detected by gravitational microlensing
arise stochastically from the underlying population of stars that intervene the observed targets, with current experiments
most of the masses of the lens stars are only known up to a broad probability distribution, although the mass of the lens star is frequently known for events in which planetary signals have been detected \citep{Gould:abundance}.
The lack of information about the planet's host star is troublesome,
since it is important to distinguish planet population statistics in the range of stellar
masses between 0.1 and 0.8~$M_\odot$, covered by microlensing, given
that current planet-formation models predict substantial differences,
in particular for the abundance of gas-giant planets \citep{IdaLin}. 

For the rather small planet samples acquired so far,
let us however neglect this issue for the time being, and just compare
planet abundance estimates that refer to the sample of probed lens stars.
Recently, there have been some discussions about planetary mass functions
that can be extracted from microlensing observations. While the discussion
by \citet{Sumi:planet} is not based on a well-defined criterion for selecting  
the considered 10~planet detections from the so far published 24 candidates towards the Galactic bulge \citep{Dominik:review}, and moreover no relation has been
given between these `detections' and the efficiency of the full observing campaigns, \citet{Gould:abundance} in contrast adopted selection criteria that lead to a well-defined event sample, and evaluated the detection efficiencies properly. However, they refer to a planetary mass function described by means of the planet-to-star mass ratio, whose value is highly questionable, given that rather obviously one does not expect the same number of half-massive planets to form around half-massive stars. In particular, coagulation and accretion processes depend on the masses of the bodies involved and their spatial density, but not on the mass of the star. Nevertheless, the sample drawn by \citet{Gould:abundance} allows for an insightful further look.

\citet{Gould:abundance} refer to 13 events with a peak magnification
$A_0 > 200$, densely monitored by MicroFUN\footnote{http://www.astronomy.ohio-state.edu/\~{}microfun/} (and other campaigns) from 2005 to 2008. Amongst those events, 2 provided a signal that indicates the
presence of a massive gas-giant planet above 150~$M_\oplus$ ($0.5~M_\rmn{jup}$), namely OGLE-2006-BLG-109 and MOA-2007-BLG-400. For such planets, the detection efficiency for orbital separations
that correspond to the `lensing zone'\footnote{Angular planet-star
separations $0.618\,\theta_\rmn{E} \leq \theta_\rmn{p} \leq 1.618\,\theta_\rmn{E}$, where $\theta_\rmn{E} = \{[(4GM)/c^2]\,(D_\rmn{L}^{-1}-D_\rmn{S}^{-1})\}^{1/2}$,
with $D_\rmn{L}$ and $D_\rmn{S}$ the distances of the lens and 
source star from the observer, respectively,
typically a range of 1.5 to 4~au.} can broadly be assumed to be of the order of
100 per cent \citep{GS98}. One would therefore estimate the abundance of such planets to be
about 15 per cent.

Rather than just focussing on events with large peak magnifications,
the PLANET collaboration \citep{PLANET:first,PLANET:EGS} has acquired data on a much larger sample of about 50 events per year with $A_0 > 2$ from 2002 to 2007, with sampling intervals of around 2~hrs or better, where the average detection efficiency for Jupiter-mass planets in the `lensing zone' for such a sample is about 15 to 20 per cent \citep{GL92}.
Only one respective planet has been reported: OGLE-2005-BLG-071Lb \citep{OB71,OB71:Dong}, as compared to expected 45--60 if those reside around each of the lens stars. This gives a rough abundance estimate of 1.5--2 per cent, 
which looks substantially smaller than what one guesses from the densely monitored events with $A_0 > 200$. The PLANET team earlier claimed
an upper abundance limit (at 95 per cent confidence) of 33 per cent on Jupiter-mass planets in the
same orbital range based on the absence of any detection amongst 42
events well-covered from 1995 to 1999
\citep{PLANET:fiveshort,PLANET:fivelong}.

The results based on the MicroFUN and PLANET data however appear to be statistically compatible, and one finds that the small-number statistics
imply large uncertainties. In fact, an Agresti-Coull confidence interval
for the planetary abundance based
on the underlying binomial distribution \citep{AgrestiCoull} at 95 per cent probability extends from 3 per cent to 43 per cent for the MicroFUN result,
and from less than 0.01 per cent to 10 or 13 per cent for the PLANET result. This also gives some indication of the unquantified uncertainties
of the statistical results quoted by \cite{Gould:abundance}.

Apart from the binomial statistics, there are some systematic uncertainties. One might in fact wonder whether there are further planets just waiting to be detected in the PLANET data that have not been spotted yet due to absence of a comprehensive systematic analysis. 
This may not be too unlikely, given that e.g.\ the event OGLE-2008-BLG-513 was initially considered to be due to a stellar binary, before a planetary
model had been suggested \citep{Gould:abundance}.
On the other hand, \citet{puzzle} have pointed to a puzzle regarding
the properties of the high-magnification events that casts doubt on whether we really understand the mechanism responsible for producing these.
Namely, a highly statistically significant correlation has been found
between the event peak magnification and the metallicity of the observed
source star (i.e.\ not the planet's host star). This sample bias is so far
not understood.

If the planetary abundance turns out to be small, its determination 
becomes more difficult. If one focuses on high-magnification peaks,
for an abundance 15 per cent, 5 per cent, or 2 per cent, the monitoring of 90, 320, or 800 microlensing events, respectively, 
would be required in order to make the half-width of a symmetric 95 per cent confidence interval match half the abundance. With an average detection efficiency of 15--20 per cent
for less favourable, but useful ($A_0 \leq 2$), hourly-sampled events 
one would require 5--6 times as many events, but given that the
number of events with $A_0 > 200$ is about 100 times smaller, such a strategy looks more feasible, in particular since with the current
detection rate of the microlensing surveys, the monitoring of $\sim\,200$ suitable events per year is possible.

While the detection efficiency for Jupiter-mass planets is a rather robust number, the sensitivity of microlensing campaigns to planets between
1~and 10~$M_{\oplus}$ (``Super-Earths'') is a strong function of planet mass and orbital separation. Therefore, the interpretation becomes substantially dependent on the choice of the considered region of planet parameter parameter space and the adopted averaging. The sparcity of data makes a meaningful assessment quite difficult. The least massive planet in
the sample adopted by \citet{Gould:abundance}, and the only one
below 50~$M_{\oplus}$, was found to have a mass of 13~$M_{\oplus}$ with a 
substantial uncertainty, so that it may or may not fall into the Super-Earth mass regime. Moreover, detection efficiencies in this region vary substantially amongst the 13 events that comprise the sample with a prominent peak close to the angular separation of the planet being equal to the angular Einstein radius of its host star. 
The situation is slightly better for the results of the PLANET campaign: 
OGLE-2005-BLG-390Lb has been estimated to have a mass between 3~and 10~$M_{\oplus}$ with a probability of 68 per cent \citep{PLANET:planet,Do:Estimate2}, and one is within the right order of magnitude by assuming a detection efficiency of about 1--2 per cent
for `lensing zone' planets \citep[c.f.][]{Bennett:Earth,390further} for an average over 300 observed events. One detection then leads to an maximum-likelihood point estimate for the respective abundance of 17--33 per cent, about 10--20 times larger than that obtained for planets above
0.5~$M_\rmn{jup}$ from the same set of observations. However, none of the derived estimates should be considered to be correct to within less than a factor 3--4, but on the other hand, they are not substantially worse either. While microlensing observations show that it is implausible that low-mass planets are rare, drawing firm conclusions is prevented by the current low-number statistics. In particular, \citet{Gould:abundance} found that their sample size is insufficient for reliably determining a power-law index in the mass function with respect to the planet-to-star mass ratio.
However, they estimate the local planet number density in orbital separation $d = \theta_\rmn{p}/\theta_\rmn{E}$ and mass ratio $q$ at
$d = 1$ and $q = 5 \times 10^{-4}$ to be $0.36$ per decade in each of the quantities. With power laws considered to range from $\propto q^{-0.2}$ to $\propto q^{-0.6}$, and
assuming a 'typical' stellar mass of $0.3~M_{\odot}$, one finds for the middle of the decade from $1~M_{\oplus}$ to $10~M_{\oplus}$, roughly 
at $m_\rmn{p} = 3~M_{\oplus}$, local number densities of $0.6$ or $1.9$, respectively. For comparison, let us account for the fact that the lensing zone covers 0.42 decades, and therefore multiply the derived abundance of 17--33 per cent by 2.4, which results in values $0.4$--$0.8$, not much different from the result found by \citet{Gould:abundance},
where one also needs to consider that the local density is not equal to the average density in the considered surrounding region. Applying the same procedure for transferring the local number density according to \citet{Gould:abundance} to a planet mass $m_\rmn{p} = 1~M_{\rmn jup}$ yields values of $0.25$ or $0.12$, for the two different power-law indices, respectively. The range of power laws gives abundance density ratios between $m_\rmn{p} = 3~M_{\oplus}$ and $m_\rmn{p} = 1~M_{\rmn jup}$ in the range from $2.5$ to $16$ (as compared to 10--20 guessed from PLANET observations).



\section{Summary and outlook}
\label{sec:outlook}

A planetary mass function is not the extension of the stellar mass function to lower masses, given that planets orbit stars. Instead, planetary mass functions more or less strongly depend on the characteristic proper properties of the respective host stars, such
as the stellar mass $M_\star$, metallicity $Z$, age $\tau$, and spin rate $\Omega$ as well. As long as the planet mass $m_\rmn{p}$, planet radius $r_\rmn{p}$, orbital semi-major axis $a$, and orbital eccentricity 
$\varepsilon$ are considered as descriptive parameters, all population statistics can be derived from three fundamental functions, namely the differential mass-radius-orbit function $\varphi(m_\rmn{p},r_\rmn{p},a,\varepsilon; M_\star, Z, \tau,\Omega)$, the fraction of stars with planetary systems $f_\mathrm{p}(M_\star, Z, \tau,\Omega)$, and the differential stellar mass function $\xi(M_\star,Z,\tau,\Omega)$.
In principle, tne fragmentation of the planetary system into $k$ planetary bodies gives us an infinite number of multiplicity indices $\zeta_k$ that correspond to the fraction of planetary systems with exactly $k$ planets, as well as respective specific mass-radius-orbit functions $\hat{\varphi}_k(m_\rmn{p},r_\rmn{p},a,\varepsilon; M_\star, Z, \tau,\Omega)$. Adopting a functional dependence on $k$ however allows for a description with a finite number of parameters. A first-order step would be to distinguish single-planet and multiple-planet systems. However, the determination of their respective fraction for all stars that host planets requires a proper assessment of the constraints on the presence of further planets in the studied systems. 

It cannot be stressed enough that the formation and evolution of planets is a crucial step towards the development of life, but it will {\em not} be understood by focusing the interest on habitable planets, rather than embracing planet populations in their amazing diversity. Moreover, rather than just optimizing planet searches for a large detection rate, the critical design feature is to follow well-defined monitoring and detection criteria that allow to carry out simulations, so that meaningful statistics can be derived.

With the probabilistic nature of the alignment of stars suitable to provide gravitational microlensing events and similarly for a planet to reveal its presence around the foreground (generally unobserved) lens star, studying planet populations by microlensing prominently comes will all difficulties and challenges that one might encounter. All conclusions that can be drawn to date not only suffer from small-number statistics, but moreover from difficulties in the assessment of the detection efficiency and the related need to refer to strictly deterministic procedures for the monitoring strategy \citep[e.g.][]{MiNDSTEp2008,Gould:abundance}. \citet{Gould:abundance} have recently presented a first statistically meaningful, but very small, sample of 13 events comprising only those events for which  source and lens stars are so closely aligned to yield a peak magnification $A_0 \geq 200$. While some fundamental statistics can be derived to within a factor 3--4, current data do not allow to do substantially better. The fact that a maximum-likelihood point estimate for the abundance of planets above $0.5~M_{\rmn jup}$ from this sample comes out as 10 times larger than what one would guess from 6 years of PLANET observations that include events with smaller peak magnifications as well, while both values are statistically compatible, demonstrates the current uncertainties. Moreover, one might wonder whether any systematics that are not fully understood affect this outcome. The relatively large number of gas-giant planets
in the sample adopted by \citet{Gould:abundance} as compared to the PLANET observations with an about 3--5 times {\em larger} total detection efficiency is somewhat surprising, in particular given that such planets are readily detected at high efficiency already for $A_0 \geq 10$ \citep{GS98}. In fact, for measuring planetary abundances, focussing on the few events with very large $A_0$ does not appear to be a promising strategy, due to the rarity of such events.


The prospects for obtaining a measurement of the planetary abundance with a desired accuracy strongly depend on the abundance itself, given that a small abundance will imply a small number of detections. This is serious limiting factor for any planet detection campaign, and the measurement
of the abundance of `second Earths' by NASA's Kepler mission is not immune to the problem of small-number statistics either. Rather than trying to estimate an abundance from a small region near the sensitivity limit, a more robust estimate for the abundance of habitable planets would arise from an interpolation between hotter and cooler planets, making use of a larger number of detected objects, and assuming that planet formation
will not be radically different just for the habitable zone.
The improvement of the statistics by a more powerful mission
such as PLATO (PLAnetary Transits and Oscillations)\footnote{{\tt http://sci.esa.int/plato}} \citep{PLATO} appears to be much desired.

The dependency of the planet population statistics on the properties of the host star implies that a quite substantial amount of data will need to be collected for properly measuring the planetary mass function and even more for determining the full mass-radius-orbit function. With the 450 reported planets so far, we can now assess 1- and 2-parameter functions (if and only if we understand any detection bias), but it requires a much larger detection rate than the recent 150 planets per year in order to understand the distribution of planets in the Universe. From 1995 to 2009, the respective time interval of acquiring 150 planets has been cut by a factor three twice (10 years, 3 years, and then 1 year), so that we see a substantial acceleration, and the Kepler mission is already contributing to boosting the planet detection rate further. Nevertheless, drawing pictures of planet parameter space in its full complexity calls for samples that are 2--4 orders of magnitude larger than those we have now. 
Looking back at the history of exoplanet detections however tells us something else: we gained a lot of insight from a few individual detections that came as a surprise and challenged the prevailing understanding. Shouldn't we expect to be surprised again when embarking on exploring further uncharted territory? Our understanding of planetary formation and evolution is not probed uniformly by a planetary mass-radius-orbit function, but certain regions of planet parameter space may prove more critical in the power to discriminate between alternative theories or to measure crucial parameters. Therefore, efficiency could be gained from observational campaigns delivering specific characteristic statistics that can be robustly determined from rather small samples.

\section*{Acknowledgments}
I would like to thank Andrew Collier Cameron, Markus Hundertmark, Christine Liebig, Sohrab Rahvar, and Yiannis Tsapras for helpful comments on the manuscript, and Michel Mayor for advice on some relevant literature.

\bibliographystyle{mn2e}
\bibliography{planetMF}

\end{document}